\documentclass[12pt]{article}
\usepackage{a4}    
\usepackage{amsmath,amssymb}

\begin{document}
\title{\bf Motion on the $n$-dimensional ellipsoid under the influence of a
harmonic force revisited
} \author{P. Di\c t\u a\\
Institute of Physics and Nuclear Engineering,\\
P.O. Box MG6, Bucharest, Romania}
\date{}
\maketitle
\footnotetext{\noindent Electronic mail: dita@hera.theory.nipne.ro}

\begin{abstract}

The $n$  integrals in involution for the  motion on the $n$-dimensional ellipsoid
under the influence of a harmonic force are explicitly  found. The classical separation of
variables is given by the inverse momentum map. In the quantum case the
Schr\"odinger equation separates into one-dimensional equations that coincide
with those obtained from the classical separation of variables. We  show
that there is a more general orthogonal parametrisation of Jacobi type that depends on
two arbitrary real  parameters. Also if there is a certain relation between the spring constants and the ellipsoid semiaxes the motion under the influence of such a harmonic potential is equivalent to the free motion on the ellipsoid.
    \end{abstract}


\section{Introduction}

In this paper we are concerned with the motion on a $n$-dimensional ellipsoid
under the influence of a harmonic potential. The problem was first posed by
Jacobi in the nineteenth century \cite{Ja} in the context of explicitly
solvable by quadratures differential equations most of them originating in completely 
integrable Hamiltonian systems. In the last decades the problem was
considered mostly by mathematicians obtaining new results including a
description of integrals in involution and connection with hyperelliptic
curves of genus $g=n$, see \cite{MH}-\cite{ Mu}. However in this approach no
explicit separation of variables was found. Quite recently we have  
obtained a new form for the $n$ prime integrals in involution for the free motion on the ellipsoid in the Jacobi
parametrisation, form  which is
very convenient for proving the separation of variables and solving the
Hamilton-Jacobi equation \cite{Di}.

The purpose of this paper is to extend our previous results in two directions:
to find a generalisation of the Jacobi orthogonal parametrisation and the most
general form of the harmonic potential for which the motion is still completely
integrable.
As  unexpected results we obtained a new orthogonal parametrisation that
depends on two arbitrary real parameters generalizing the Jacobi one, and  
if the spring constants $k_1$ and $k_2$ along two axes of symmetry satisfy the
relation $a_1\,k_1=a_2\,k_2$, where $a_1$ and $a_2$ are the squares of the
corresponding semiaxes then the motion under the influence of such a harmonic
potential is the free motion on the ellipsoid.
In our approach the separation of variables is realized by the inverse of the
momentum map that provides an explicit factorisation into Liouville tori. In
the standard approach the separation of variables is a rather difficult
problem, see e.g. \cite{KKM} where the separated variables are defined as
zeros of the off diagonal elements of the associated  Lax matrix.

In Section 2 we define a two-dimensional family of Jacobi elliptic coordinates
and in Section 3 we prove explicitly their orthogonality. The integrals in
involution for the free motion on the ellipsoid are found in Section 4 and we
show that they are a particular case of a more general form. The most general
form of the harmonic potential for which the motion is still integrable is
given in Section 5 and the classical separation of variables is obtained in
Section 6. The separation of the Schr\"odinger equation is shown in Section 7
and the paper ends with Conclusion. In Appendix are collected a few known
properties of the Vandermonde determinant which are used in the paper.

\section{Generalized Jacobi's coordinates on the\\
 $n$-dimensional ellipsoid}

Usually the $n$-dimensional ellipsoid is viewed as a surface immersed in a
$n+1$-dimensional Euclidean space, defined by the equation
$${x_1^2\over a_1}+{x_2^2\over a_2}+\dots+{x_{n+1}^2\over a_{n+1}}=1$$
where $a_i,\, i=1,\dots, n+1$ are positive numbers, $a_i\in {\bf R_+}$, and
we suppose that they are  ordered such that  $a_1>\dots>a_{n+1}> 0$. We describe the
$n$-dimensional ellipsoid by intrinsic coordinates and here we slightly
generalize the Jacobi's approach by defining the generalized elliptic
coordinates as the solutions of the equation 
$$\sum_{i=1}^{n+1}\, {b_i\,x_i^2\over a_i(b_i-\lambda)}=1\eqno(1)$$
with $b_i\in {\bf R^*}$,  $b_i\ne b_j$, for $i\ne j$, $i,j=1,\dots,n+1$.
We rewrite (1) in the form
$$1+\sum_{i=1}^{n+1}{b_i\,x_i^2\over
a_i(\lambda -b_i )}={\lambda\,Q(\lambda)\over P(\lambda)}\eqno(2)$$
where $P(\lambda)$ and $Q(\lambda)$ are the monic polynomials

$$Q(\lambda)=\prod_{i=1}^n \,(\lambda-u_i),\quad
P(\lambda)=\prod_{i=1}^{n+1}\,(\lambda-b_i)$$
Here $u_i,\,i=1,\dots,n$ are the Jacobi coordinates and the factor $\lambda$
comes from the fact  that $\lambda=0$ is a solution of (1) when the coordinates
$x_i$ lie on the ellipsoid.

Calculating the residues on both left and right sides of equation (2) one gets
$$x_i^2=a_i\,{Q(b_i)\over P'(b_i)},\,\, i=1,\dots,n+1\eqno(3)$$
If $b_i=a_i,\,\, i=1,\dots,n+1$, (3) gives the usual Jacobi parametrization
of the $n$-dimensional ellipsoid. For $a_i=r^2,\, i=1,\dots,n+1$, the relation (3)
gives the orthogonal parametrization of the $n$-dimensional sphere of radius
$r$ and in this case (1) has the form
$$\sum_{i=1}^{n+1}{x_i^2\over \lambda-b_i}=0$$
Our approach is sufficiently general and gives the parametrization of any
quadric defined by $a_i \in{\bf R}^*$.

The parametrisation (3) depends on $2n+2$ parameters, half of them $b_i\in
{\bf R^*},\, \, i=1,\dots,n+1$, being arbitrary real numbers and for this reason
the corresponding Jacobi coordinates $u_i$ could not be  orthogonal. However it
does exist an orthogonal parametrization, different from the usual one, which
depends on $n+3$ parameters. In other words there is more than one orthogonal
system of coordinates of Jacobi type and this situation could be interpreted as
a hidden symmetry of the problem. We found that this symmetry is
two-dimensional and the independent parameters in equation (3) leading to orthogonal
coordinates may be chosen  as $a_i,\, i=1,\dots,n+1, b_1$ and $b_2$, the last
two being arbitrary non-zero  real numbers with $b_1\ne b_2$.

To find the new parametrization we deduce from (3)

$${\dot{x_i}\over x_i}={1\over 2}\sum_{l=1}^n\,{\dot{u_l}\over u_l-b_i}$$
and $$\sum_{i=1}^{n+1}\,\dot{x}_i^2={1\over 4}\sum_{l,m=1}^n\,
\dot{u}_l \,\dot{u}_m\,\sum_{i=1}^{n+1}\,{x_i^2\over (u_l-b_i)(u_m-b_i)}$$
Taking into account the relation (3), a careful inspection shows that the function 
$$c_{lm}= \sum_{i=1}^{n+1}\,{x_i^2\over (u_l-b_i)(u_m-b_i)},\quad l\ne m$$
is a symmetric polynomial with respect to $u_i$ in its $n-2$ variables, having degree
equal to  $n-2$ and $n-1$ independent coefficients. An important property is
that these coefficients do not depend on $l$ and $m$, i.e.  for all $l\ne m$,
$c_{lm}$ defines a single function; instead of $n(n-1)/2$ different
polynomials we have only one. Imposing now the condition $c_{lm}\equiv 0$ for
$l\ne m$ we obtain $n-1$   equations which can be solved with respect to
$b_i$. So the number of parameters of the new parametrization is $2
n+2-(n-1)=n+3$.

To see what happens let consider the case $n=2$; in this case the condition
$c_{12}\equiv 0$ is equivalent to the relation
$$b_3={a_1b_2-a_2b_1+a_3(b_1-b_2)\over a_1-a_2}$$
By iteration we obtain for $n=4$ the additional constraint
$$b_4={a_1b_2-a_2b_1+a_4(b_1-b_2)\over a_1-a_2}$$
Thus we suppose that the general case is given by
$$b_n={a_1b_2-a_2b_1+a_n(b_1-b_2)\over a_1-a_2},\,\, n=3,\dots, n+1\eqno(4)$$
and in the next Section we prove that the parametrisation (3) with $b_i$,
$i=3,\dots, n+1$, given by (4) and $b_1$ and $b_2$, $b_1\ne b_2$, arbitrary
real non-vanishing  numbers furnishes $n$ orthogonal elliptic coorddinates.
Before doing that we introduce a more uniform notation by defining two new
parameters
$$\alpha={a_1\,b_2- a_2\,b_1\over a_1- a_2}, \quad {\rm and  }\quad
\beta={b_1-b_2\over a_1- a_2}$$
Solving with respect to $b_1$ and $b_2$ we find
$$b_i=\alpha +\beta\,a_i,\,\,i=1,\dots, n+1\eqno(4')$$
In the next Section we prove that the two-dimensional family of Jacobi
coordinates whose parametrisation is given by the relation (3) with $b_i$ as in
(4$'$) is orthogonal for all real values of $\alpha\in {\bf R}$ and $\beta\in {\bf R^*}$.

\section{Orthogonality property} 

For proving the orthogonality of the new coordinates we write  equation (2) in
another form
$$\sum_{i=1}^{n+1}\,\frac{x_i^2}{a_i(\lambda -b_i )}
=\frac{Q(\lambda)}{P(\lambda)}$$ In the last relation we substitute
$b_i=\alpha+\beta\,a_i$, $i=1,\dots,n+1$, and after minor transformation we get
$$\sum_{i=1}^{n+1}\,\frac{x_i^2}{\alpha+\beta\,a_i -\lambda}
=\frac{1}{\beta}\left(1-\frac{(\lambda-\alpha)Q(\lambda)}{P(\lambda)}
\right)\eqno(5)$$ For $l\ne m$ we have
$$c_{lm}=\sum_{i=1}^{n+1}\frac{x_i^2}{(u_l-b_i)(u_m-b_i)}=
\frac{1}{u_l-u_m}\left(\sum_{i=1}^{n+1}\frac{x_i^2}{u_m-b_i}-\sum_{i=1}^{n+1}\frac{x_i^2}{u_l-b_i}\right)$$  Substituting  $b_i=\alpha+\beta\, a_i$ in the last relation and   using (5) and the property $Q(u_i)=0,\, i=1,\dots,n$, we find that $c_{lm}=0$ for $l\ne m$, i.e. the new coordinates are orthogonal, and our assumption expressed by the relation (4) is true.

For $l=m$ we have $$c_{ll}=\sum_{i=1}^{n+1}\frac{x_i^2}{(u_l-b_i)^2}=-\left.\frac{d}{dz}\sum_{i=1}^{n+1}\frac{x_i^2}{(z-b_i)}\right|_{z=u_l}\eqno(6)$$
and we need a calculation of the last sum. It can be obtained by differentiating  (5)
 with respect to $\lambda$. We use   this  result  to write the Lagrangean in the form
$${\cal L}=\frac{1}{2}\sum_{i=1}^{n+1}{\dot{x}_i^2}=\sum_{j=1}^{n}g_j\dot{u}_j^2\eqno(7)$$
where $g_j=-\frac{1}{4\,\beta}\frac{(u_j-\alpha)Q'(u_j)}{P(u_j)}$ and $ Q'(u_j)=d\,Q(x)/d\,x|_{x=u_j}$. For $\alpha=0$ and $\beta=1$ one recovers the usual result \cite{Mo}.

Following the standard procedure we find the Hamiltonian of the free motion on the $n$-dimensional  ellipsoid
$${\cal H}=\sum_{j=1}^{n}\,p_j\,u_j-{\cal L}=-2\,\beta\sum_{j=1}^{n}\,g^i\,p_i^2\eqno(8)$$
where $g^i=P(u_i)/(u_i-\alpha)\,Q'(u_i)$ and $u_j,\,p_j$ are canonical coordinates.

Unlike the classical result we have obtained that the Hamiltonian of the geodesic motion is not uniquely defined, it depends continuously and non-trivially on two arbitrary real  parameters $\alpha$ ans $\beta$. By changing these parameters one changes the classical state of the system if the latter one is defined as a point in the phase space but the form of the energy does not change. This property could be interpreted as a gauge symmetry of the classical Jacobi problem. Under the transformations
$$x_i^2\,\rightarrow \,a_i\,\frac{Q(b_i)}{P'(b_i)}\,,\quad b_i\rightarrow \alpha+\beta\, a_i,\quad\alpha\in {\bf R},\quad\beta\in {\bf R^*}$$
the Lagrangean ${\cal L}$ and the Hamiltonian  ${\cal H}$ remain invariant, and {\it ipso facto} the equations of motion. Since with given initial conditions the physical motion is unique the only freedom we have in proving the uniqueness is the reparametrization of time. Generally speaking this fact rises a new problem, namely that of  finding  all the metrics which lead to the same physical motion; in other words how large  
the hidden symmetry is, or how many geometries describe the same physical process. A solution to this problem could be of interest in the study of more complicated models arising in classical and quantized field theories.

\section{Integrals in involution}

We introduce now $n$ integrals in involution that are linear independent. With this aim we define the symmetric functions of the polynomials $Q'(u_j)\equiv Q_j(u_j)$
$$ Q_j(u_j)=\sum_{k=0}^{n-1}\, u_j^k\,S_{n-k-1}^{(j)}\eqno(9)$$
where $S_k^{(j)}=(-1)^k\,\sigma_k(u_1,\dots,u_{j-1},u_{j+1},\dots,u_n)$ and $\sigma_k$ is the symmetric polynomial of degree $k$. By construction the coordinate $u_j$ does not enter the symmetric sum  $S_{k}^{(j)}$, $k=0,1,\dots, n-1$. The following functions 
$$H_k=\sum_{l=1}^n\,S_{k-1}^{(l)}\,g^l\,p_l^2,\quad k=1,\dots,n\eqno(10)$$
with $H_1=-{\cal H}$ up to a numerical factor are $n$ integrals in involution for the geodesic motion on the ellipsoid.

The integrals of motion $H_k$, $k=1,\dots,n$ in the above form were found by us in ref. \cite{Di}; for another approach see e.g. Moser \cite{MH}.

An inspection of (10) shows that for each degree of freedom the contribution to the Hamiltonian $H_k$ is given by  a product of two factors. The first one, the ``kinematical'' factor, depends on a special structure, in our case the Vandermonde structure defined by the ratio $f_1(u_1,\dots,u_n)=S_{k-1}^{(i)}/Q'(u_i)$, and the second one, the kinetic energy, $ f_2(u_i,p_i)=p_i^2\,P(u_i)/(u_i-\alpha)$ depends on the ``physics'', in our case the geometry of the body. The important issue is the factorizatin $f_1(u_1,\dots,u_n)\cdot f_2(u_i,p_i)$ where $f_1$ has no momentum dependence and $f_2$ depends only on a pair of canonical variables and nothing else.

Let $g(p,u)={\cal H}(p,u)$ be an arbitrary function depending on the canonical variables $p$ and $u$ which is invertible with respect to the momentum $p$. As we will see later  
the invertibility condition is necessary for the separation of variables in the Hamilton-Jacobi equation. In particular we may suppose that ${\cal H}(p,u)$ is a one-dimensional Hamiltonian. For each $n\in {\bf N}$ we define an $n$-dimensional integrable model by giving its $n$ integrals in involution
$${\bf H}_k(p,u)=\sum_{i=1}^n\,\frac{S_{k-1}^{(i)}}{Q'(u_i)}\,g(p_i,u_i),\quad k=1,\dots,n \eqno(10')$$
We will prove the involutive property in the more general case (10$'$). We have
$$\{{\bf H}_k,{\bf H}_l\}=\sum_{j=1}^n\left(\frac{\partial {\bf H}_k}{\partial u_j}\,\frac{\partial {\bf H}_l}{\partial p_j}-\frac{\partial {\bf H}_k}{\partial p_j}\,\frac{\partial {\bf H}_l}{\partial u_j}\right)=\sum_{j=1}^n\frac{\partial g(p_j,u_j)}{\partial p_j}\left(\frac{S_{l-1}^{(j)}}{Q'(u_j)}\,\frac{\partial{\bf H}_k}{\partial u_j}-\frac{S_{k-1}^{(j)}}{Q'(u_j)}\,\frac{\partial{\bf H}_l}{\partial u_j}\right)=$$

$$\sum_{j=1}^n\sum_{i=1}^n\frac{1}{Q'(u_j)}\frac{\partial g(p_j,u_j)}{\partial p_j}\left(S_{l-1}^{(j)}\frac{\partial}{\partial u_j}\left( g(p_i,u_i)\frac{S_{k-1}^{(i)}}{Q'(u_i)}\right)-S_{k-1}^{(j)}\frac{\partial}{\partial u_j}\left( g(p_i,u_i)\frac{S_{l-1}^{(i)}}{Q'(u_i)}\right)\right)=$$
$$\sum_{j=1}^n\sum_{i=1}^n\frac{1}{Q'(u_j)}\frac{\partial g(p_j,u_j)}{\partial p_j}\frac{\partial}{\partial u_j}\left(\frac{g(p_i,u_i)(S_{k-1}^{(i)}S_{l-1}^{(j)} -S_{k-1}^{(j)}S_{l-1}^{(i)})}{Q'(u_i)}\right)$$ 
The last step was possible because the symmetric functions $S_k^{(j)}$ and  $S_l^{(j)}$ do not depend on $u_j$. Looking at the last expression it is easily seen that the partial derivative with respect to $u_j$ vanishes for $i=j$. For  $i\ne j$ we have to show that
$$\frac{\partial}{\partial u_j}\frac{S_{k-1}^{(i)}S_{l-1}^{(j)} -S_{k-1}^{(j)}S_{l-1}^{(i)}}{u_i-u_j}= 0$$
but this is a consequence of the following identities
$$\frac{\partial}{\partial u_j}\,S_{k-1}^{(i)}=-S_{k-2}^{(i,j)}\quad{\rm and}\quad S_{k-1}^{(i)}- S_{k-1}^{(j)}=(u_i-u_j)S_{k-2}^{(i,j)}$$
where the upper index $(i,j)$ means that the corresponding expression does not depend on both $u_i$ and $u_j$. In this way we have shown that $\{{\bf H}_k,{\bf H}_l\}=0$

\section{Harmonic potential}

In the following we want to find the most general form of the harmonic potential for which the motion on the ellipsoid under the influence of this potential is still 
integrable. Moser says that ``the motion on an ellipsoid under the influence of a potential $|x|^2$ is also integrable [and] this was shown already by Jacobi \cite{MH}''. On the other hand Arnold makes a stroger statement: ``Jacobi showed that the problem of free motion on an ellipsoid remains integrable if the point is subjected to the action of an elastic force whose direction passes through the center of the ellipsoid \cite{AKN}'', which might be understood as suggesting that the spring constants on different axes  are different. We start with  the most general form for the harmonic potential
$${\cal U}=\frac{1}{2}\sum_{i=1}^{n+1}k_i\,x_i^2$$
with $k_i\ne k_j$ for $i\ne j$ and we look for conditions on $k_i$ for which the motion is integrable. We show that the most general form for ${\cal U}$ depends on the semiaxes of the ellipsoid and two arbitrary parameters which can be taken $k_1$ and $k_2$.
By substitution of the relation (3) in the above formula we get
$$ {\cal U}=\frac{1}{2}\sum_{i=1}^{n+1}k_i\,x_i^2=\sum_{k=0}^n\left(\sum_{i=1}^{n+1}\frac{k_i\,a_i}{P'(b_i)}b_i^k\right)S_{n-k}$$
where $S_k$ are up to a $\pm$ sign the symmetric functions 
of $u_1,\dots,u_n$. This means that $ {\cal U}$ is a  polynomial of degree $n$
in the elliptic coordinates defined by the equations (3)-(4$'$). The motion described by the Hamiltonian ${\it H}={\cal H}+{\cal U}$ is  completely integrable iff the coefficients of all the products $u_i\cdot\cdot\cdot u_l$ vanish such that ${\cal U}$ in the new variables should have the form ${\cal U}=A+B(u_1+\dots+u_n)$.
The vanishing of these coefficients leads to the relations
$$k_i=\frac{a_1a_2(k_1-k_2)-a_i(a_1k_1-a_2k_2)}{a_i(a_2-a_1)},\,\, i=3,4,\dots,n+1$$
which shows that  the spring constants do not depend  on the previous introduced parameters $\alpha$ and $\beta$. Like the preceding case we define two new parameters
$$\gamma=\frac{a_1a_2(k_1-k_2)}{a_2-a_1}\quad {\rm and}\quad \delta=\frac{a_2k_2-a_1k_1}{a_2-a_1},\quad \gamma,\delta\in{\bf R}$$
such that $k_i$ has the dependence
$$k_i(a_i)=\frac{\gamma+\delta\,a_1}{a_i},\, i=1,\dots,n+1$$
that can be viewed as the action of a special element of $G\,L_2^+({\bf R})$ which transforms the right half-space into itself. For $\gamma=0$ one recovers the classical result. Thus we obtained that there is a two-dimensional family of coefficients $k_i$, which depend also on the semiaxes of the ellipsoid, for which the motion is integrable. Making all the calculation one finds
$${\cal U}=\frac{1}{2}\left(n\,\frac{\alpha\delta}{\beta}+\gamma+\delta(a_1+a_2)+\sum_{k=3}^{n+1}a_k-\frac{\delta}{\beta}\sum_{i=1}^nu_i\right)$$
which is the most general form of the harmonic potential for which the motion is completely integrable. If $\delta=0$ or $a_1k_1=a_2k_2$  the harmonic potential in elliptic coordinates reduces to a constant, i.e. the motion is the free geodesic motion,  fact which is noticed for the first time.

\section{Separation of variables}
The Hamiltonian of the problem is  $H={\cal H}+{\cal U}$ and we do not know yet how the other prime integrals look. We shall neglect the constants terms appearing in ${\cal U}$ such that
$$H=-2\beta\sum_{i=1}^{i=n}\frac{P(u_i)}{(u_i-\alpha)Q'(u_i)}\,p_i^2-\frac{\delta}{2\beta}\sum_{i=1}^n u_i\eqno(11)$$
The Hamiltonians $H_k$, see the relations (10), depend on the symmetric functions $S_k^{(j)}$. A similar sum appears also in the potential, namely $S_1=-\sum_{i=1}^n u_i$, where $S_k$ are defined similarly to equation (9) by the relation
$$Q(x)=\prod_{i=1}^n(x-u_i)=\sum_{k=0}^nx^k\,S_{n-k}\eqno(12)$$
It is easily seen that $dS_k/du_j=-S_{k-1}^{(j)}$. For the other prime integrals we define the potentials ${\cal U}_k=\frac{\delta}{2\,\beta}\,S_k$ for $k=2,\dots,n$ such that the integrals in involution are given by
$${\cal H}_k=-2\beta\sum_{i=1}^{i=n}\frac{P(u_i)}{(u_i-\alpha)Q'(u_i)}\,p_i^2 +\frac{\delta}{2\,\beta}S_k,\quad k=1,\dots,n\eqno(13)$$

The involution property follows straightforward
$$\{{\cal H}_i,{\cal H}_j\}=\{{\cal H}_i,{\cal H}_j\}+\{{\cal H}_i,{\cal U}_j\}+\{{\cal U}_i,{\cal H}_j\}+\{{\cal U}_i,{\cal U}_j\}=$$
$$\sum_{l=1}^n\left(\frac{\partial{\cal U}_i}{\partial u_l}\frac{\partial{\cal  H}_j}{\partial p_l}-\frac{\partial{\cal  H}_i}{\partial p_l}\frac{\partial{\cal U}_j}{\partial u_l}\right)=-2\delta \sum_{l=1}^n\frac{P(u_l)\,p_l}{(u_l-\alpha)Q'(u_l)}\left(S_{i-1}^{(l)}S_{j-1}^{(l)}-S_{j-1}^{(l)}S_{i-1}^{(l)}\right)=0$$
The next important point is the separaton of variables.

Let $M^{2n}\simeq T^*({\bf R}^n)$ be the canonically symplectic phase space of the dynamical system defined by the Hamilton functions (13). We define the momentum map by
$${\cal E}: M^{2n}\rightarrow {\bf R}^n :\, M_{{\bf h}}=\{(u_i,p_i): H_i=-h_i,\,\, i=1,\dots,n\},\, h_i\in {\bf R}\eqno(14)$$
This application is such that ${\cal E}^{-1}(M_{{\bf h}})$ realizes the separation of variables giving an explicit factorisation of Liouville's tori into one-dimensional ovals. Our goal is to construct explicitely the application  ${\cal E}^{-1}(M_{{\bf h}})$ and for doing that we write the system (13) in a matrix form. With the notation $f(u,p)=P(u)\,p^2/(u-\alpha)$ (13) is written as

$$ 2\beta\left(\begin{array}{cccc}
S_0^{(1)}/Q'(u_1)\,\,&S_0^{(2)}/Q'(u_2)&\dots&S_0^{(n)}/Q'(u_n)\\
S_1^{(1)}/Q'(u_1)\,\,&S_1^{(2)}/Q'(u_2)&\dots&S_1^{(n)}/Q'(u_n)\\
\dots&\dots&\dots&\dots\\
S_{n-1}^{(1)}/Q'(u_1)&S_{n-1}^{(2)}/Q'(u_2)&\dots&S_{n-1}^{(n)}/Q'(u_n)\\
\end{array}\right)\left(\begin{array}{c}
f(u_1,p_1)\\
f(u_2,p_2)\\
\dots\\

f(u_n,p_n)\\
\end{array}\right)-$$
$$\frac{\delta}{2\beta}\left( 
\begin{array}{c}
S_1\\
S_2\\
\dots\\
S_n\\
\end{array}
\right)=\left( \begin{array}{c}
h_1\\
h_2\\
\dots\\
h_n\\
\end{array}
\right)$$
Multiplying to left by the matrix

$$V=\left( \begin{array}{cccc}
u_1^{n-1}&u_1^{n-2}&\dots&1\\
u_2^{n-1}&u_2^{n-2}&\dots&1\\
\dots&\dots&\dots&\dots\\

u_n^{n-1}&u_n^{n-2}&\dots&1\\
\end{array} \right)\eqno(15)$$
we get a diagonal matrix with its non-zero elements equal to unity that multiplies the  vector column $(f(u_1,p_1),f(u_2,p_2),\dots,,f(u_n,p_n))^t$, where $t$ means transpose. In the proof one make use of the relation A2 given in Appendix. The solution is 
$$f(u_i,p_i)=-\frac{\delta}{4\beta^2}\,u_i^n+\frac{1}{2\beta}\sum_{k=1}^n\,h_k\,u_i^{n-k},\quad i=1,\dots,n\eqno(16)$$
where the first term on the right hand side is a result of the property $Q(u_i)=0$, see equation (12), written in the form $u_i^n=-\sum_{k=0}^{n-1}u_i^k\,S_{n-k}$. The above relations have the classical form \cite{Sk}
$$\varphi(x_i,p_i,h_1,\dots,h_n)=0,\quad i=1,\dots,n$$
which for $h_i=c_i,\, i=1,\dots,n$ give an explicit parametrization of Liouville's tori.

Taking into account the form of $f(u_i,p_i)$ and defining $R(u_i)=(u_i-\alpha)(-\frac{\delta}{4\beta^2}\,u_i^n+\frac{1}{2\beta}\sum_{k=1}^n\,h_k\,u_i^{n-k})$ we get
$$p_i=\epsilon_i\sqrt{\frac{R(u_i)}{P(u_i)}},\quad i=1,\dots,n$$
where $\epsilon_i=\pm 1$.
 By the substitution $p_i\rightarrow\partial S/ \partial t_i$ the
Hamilton-Jacobi equation separates and the solution has the form $$S({\bf
h},u_1,\dots,u_n)=\sum_{i=1}^n\,\epsilon_i
\int_{u_i^0}^{u_i}\sqrt{\frac{R(w)}{P(w)}}\, dw$$ On the last expression
one can see that all the subtleties  of the problem are encoded by the
hyperellitic curve $y^2=P(x)R(x)$ whose genus is $g=n$.

The motion described by the prime integrals (10$'$) is also separable and
$$g(u_i,p_i)=\sum_{k=0}^{n-1}\,h_{n-k}u_i^k,\quad i=1,\dots,n$$
For obtainaing the geodesic equations $g(u,p)$ has to be invertible with respect to $p$. With the notation ${\cal R}(u)=\sum_{k=0}^{n-1}\,h_{n-k}u_i^k$
the momentum is given by
$$p_i=g^{-1}({\cal R}(u_i))$$
where $g^{-1}$ denotes the inverse of $g$ with respect to $p$ and the solution of the Hamilton-Jacobi equation has now the form
$$S({\bf h},u_1,\dots,u_n)=\sum_{i=1}^n\,\epsilon_i \int_{u_i^0}^{u_i}g^{-1}({\cal R}(w))\, dw$$
The above formulae allow us to choose new canonical variables \cite{A}, and in the last case we may take ${\cal Q}_k={\cal H}_k $, $k=1,\dots,n$ and the canonically conjugated variables  ${\cal P}_i,\, i=1,\dots,n$. The Hamilton equations are
$$\dot{{\cal Q}}_i=0,
\,\, i=1,\dots,n$$
$$\dot{{\cal P}}_1=-1,\dot{{\cal P}}_i=0,\, i=2,\dots,n$$
and therefore ${\cal Q}_i=h_i, i=1,\dots,n$ and ${\cal P}_1=-t+g_1,\,{\cal P}_k=g_k,\, k=2,\dots,n$ with $g_i,h_i\in {\bf R}, i=1,\dots,n$. Because 
$${\cal P}_i=-\frac{\partial S}{\partial{\cal Q}_i}=-\frac{\partial S}{\partial h_i}=-\sum_{i=1}^n\int_{u_i^0}^{u_i} (g^{-1})'({\cal R}(w))w^{n-i}\,dw$$
where $(g^{-1})'(z)=dg^{-1}(z)/dz$ one obtains the system of equations
$$-t\,\delta_{1,j}+b_j=\sum_{i=1}^n\int_{u_j^0}^{u_j} (g^{-1})'({\cal R}(t))t^{n-j}\,dt,\quad j=1,\dots,n$$
which represents the implicit form of the geodesics, and shows that the canonical equations are integrable by quadratures. In the above formulae we singled out the first prime integral ${\cal H}_1$; if we start with ${\cal H}_k$ as Hamiltonian  then the change in the above formulae is $\delta_{1j}\rightarrow \delta_{kj}$

\section{Quantisation}

For the beginning we consider the Hamiltonian ${\cal H}_1$ given by equation (13). It is well known that because of the ambiguities concerning the ordering of $u$ and $p$  we must use the Laplace-Beltrami operator \cite{Po}. Its general form is $\Delta=\frac{1}{\sqrt{g}}(\sqrt{g}g^{ij}p_j),\, i,j=1,\dots,n$, where $g=det(g_{ij})$and $g_{ij}$ is the metric tensor. In our case $g_{ij}=-\frac{1}{4\beta}\frac{(u_i-\alpha)Q'(u_i)}{P(u_i)}\delta_{ij}$ and $g^{ii}=P(u_i)/(u_i-\alpha)Q'(u_i)$. Let $V_n$ denote the determinant of V, Eq.(15), and $V_n^{(j)}$ the determinant of  the matrix obtained from $V$ by removing the last row and the $j$-th column. Using the relations A.3 and A.4 from Appendix we find that up to an inessential numerical factor
$$g=V_n^2\,\prod_{i=1}^n\frac{u_i-\alpha}{P(u_i)}$$
and using it the Schr\"odinger equation generated by ${\cal H}_1$ is written, after some simplification, in the form
$$2\beta\sum_{i=1}^n\frac{1}{V_n}\,\sqrt{\frac{P(u_i)}{u_i-\alpha}}\,\frac{\partial}{\partial u_i}\left((-1)^{n-i}V_{n-1}^{(i)}\,\sqrt{\frac{P(u_i)}{u_i-\alpha}}\,\frac{\partial\Psi}{\partial u_i} \right)-\left(\frac{\delta}{2\beta} \sum_{i=1}^n u_i\right)\Psi=E_1\Psi$$
Since the factor $V_{n-1}^{(i)}$ does not depend on $u_i$ it can be pulled out of the bracket and the preceding equation takes the form
$$\sum_{i=1}^n\,(-1)^{n-i}V_{n-1}^{(i)}\,\sqrt{\frac{P(u_i)}{u_i-\alpha}}\,\frac{\partial}{\partial u_i}\left(\sqrt{\frac{P(u_i)}{u_i-\alpha}}\,\frac{\partial\Psi}{\partial u_i} \right)-\left(\frac{\delta}{4\beta^2}V_n\sum_{i=1}^n u_i\right)\Psi=\frac{E_1}{2\beta}V_n\Psi$$
Now we make use of the Jacobi identities A.5 and find
$$\sum_{i=1}^n\,(-1)^{n-i}V_{n-1}^{(i)}\left[\sqrt{\frac{P(u_i)}{u_i-\alpha}}\,\frac{\partial}{\partial u_i}\left(\sqrt{\frac{P(u_i)}{u_i-\alpha}}\,\frac{\partial\Psi}{\partial u_i} \right)+\left(-\frac{\delta}{4\beta^2}\,u_i^n+\sum_{k=0}^{n-1}c_{n-k}u_i^k\right)\Psi\right]=0$$
which is equivalent to $n$ equations of the form
$$\sqrt{\frac{P(u_i)}{u_i-\alpha}}\,\frac{\partial}{\partial u_i}\left(\sqrt{\frac{P(u_i)}{u_i-\alpha}}\,\frac{\partial\Psi_i}{\partial u_i} \right)+\left(-\frac{\delta}{4\beta^2}\,u_i^n+\sum_{k=0}^{n-1}c_{n-k}u_i^k\right)\Psi_i=0,\quad i=1,\dots,n\eqno(17)$$
Here $c_1=-E_1/2\beta$ and the other $c_k$ are arbitrary.

The direct approach, starting with equation (16), is simpler the problem being one-dimensional and one gets the same equation (17). It has the advantage that the arbitrary coefficients $c_k$ are identified to $c_k=-h_k/2\beta$, i.e. $c_k$ are the eigenvalues of the Hamiltonians ${\cal H}_k$.

In this way the solving of the Schr\"odinger equation was reduced to the solving of a Sturm-Liouville equation whose general form is
$$-\frac{d}{dx}\left(p(x)\frac{df(x)}{dx}\right)+v(x)f(x)=\lambda r(x)f(x)$$
and the above equation has to be resolved on an interval $[a,b]$. It is well known that its eigenfunctions will live in a Hilbert space iff $p(x)r(x) > 0$ on $[a,b]$. If $p(x)$ has a continuous first derivative and $p(x)r(x)$ a continuous second derivative then by making the following coordinate and function transforms
$$\varphi =\int^u\left(\frac{r(x)}{p(x)}\right)^{1/2}\,dx, \quad \Phi=(r(u)p(u))^{1/4}f(u)\eqno(18)$$
we bring the preceding equation to the standard form
$$-\frac{d^2\Phi}{d\varphi^2}+q(\varphi)\Phi=\lambda\,\Phi$$
where
$$q(\varphi)=\frac{\mu^{''}(\varphi)}{\mu(\varphi)}-\frac{v(u)}{r(u)},\quad \mu(\varphi)=(p(u)r(u))^{1/4}$$
and $u=u(\varphi)$ is the solution of the inverse Abel problem (18).

In our case, Eq.(17), the transformation is 
$$\varphi=\int_{u_0}^u\left(\frac{R(u)}{P(u)}\right)^{1/2} du$$
and the Schr\"odinger equation gets
$$-\frac{d^2\Phi}{d\varphi^2}+\frac{\mu^{''}(\varphi)}{\mu(\varphi)}{\Phi=h_1}\Phi\eqno(19)$$
where $\mu(\varphi)=(R(u(\varphi))^{1/4}$ and in $R(u)$ we made the rescaling $h_k\rightarrow h_k/h_1, k=1,\dots,n$, i.e. the solving of (17) is equivalent to solve (19) which represents the motion of a one-dimensional particle in the potential generated by $R(u(\varphi))$.

For $n=1$ and $\delta=0$ (19) is nothing else than the equation for the one-dimensional rotator
$$\frac{d^2\Psi}{d\varphi^2}+ l^2\Psi=0$$
with the solution $\Psi(\varphi)=\frac{1}{\sqrt{2\pi}}e^{i\,l\,\varphi}, l\in Z$, etc. In all the other cases we have to make use of the theory of hyperelliptic curves and/or $\theta$-functions in order to obtain explicit solutions. This problem will be treated elsewhere.
\section{Conclusion}

In this paper we have obtained the most general form of 
the harmonic potential for which the motion of a point on the $n$-dimensional
ellipsoid  is completely integrable and we found another form of the integrals
in involution. The advantage of our approach  is that separation of variables
is very easy being an immediate consequence of the St\"ackel structure
appearing into equations of motion. We have shown that the Vandermonde structure is
enough powerful to allow construction of new $n$-dimensional completly
integrable models. Two such models could be given by the one-dimensional
Hamiltonians, $g(u,p)=(sin\,u/u)\,p^2$ and $g(u,p)=tg\,u\,e^{\alpha p}$, see
(10$'$). These models are interesting since in the first example $g(u,p)$
is a function wich have  a denumerable number of zeros and the second one  has
a denumerable number of zeros and poles, in both cases the hyperelliptic curve
being of infinite genus. Thus these models show that the dimension $n$ of the
system has no direct connection with the number of zeros and/or poles of the
function $g(u,p)$.

Other examples of $n$-dimensional Hamiltonians are 
obtained for example from the 
many-body elliptic Calogero-Moser model \cite{BMMM} 
or the elliptic Ruijenaars model \cite{Ru}, starting 
with the one-dimensional
Hamiltonians $H_{CM}(u,p)=p^2/2+\nu^2{\cal P}_{\tau}(u)$ and
$H_R(u,p)=cosh(\alpha p)\sqrt{1-2(a\nu)^2{\cal P}_{\tau}(u)}$ respectively,
where ${\cal P}_{\tau}(u)$ is the Weierstrass function and using the above machinery.

In conclusion we discovered a new method for obtaining $n$-dimensional  completely integrable systems starting with  one-dimensional Hamiltonians. Also interesting is the existence of a class of orthogonal metrics \`a la Jacobi for which the Lagrangean is gauge invariant. This result rises the problem of description of all the metrics that lead to the same physics. A step in this direction could be the revivification of the techniques discovered  by St\"ackel, Levi-Civita, Painlev\'e and many others which falled into undeserved oblivion. 
\newpage
{\bf \Large A\quad Appendix}

\vskip5mm
Herewith we collect a few known properties of the Vandermonde determinant, the novelty being their presentation in the context of algebraic duality. Let $a_i,\, i=1,\dots,n$ be $n$ real  (complex) numbers. We define the polynomial
$$P(x)=\prod_{i=1}^n(x-a_i)=\sum_{k=0}^n S_kx^{n-k}\eqno{\rm A.1}$$
where $S_k=(-1)^k\sigma_k(a_1,\dots,a_n)$ and $\sigma_k$ denotes the elementary symmetric polynomilas of degree $k$. From A.1 we have
$$P(a_i)=0, \quad i=1,\dots,n$$
i.e. the vectors $X_1=(S_0,S_1,\dots,S_n)$ and $X_2=(a^n,\dots,a,1)$ are
orthogonal $(X_1,X_2)=0$ under the usual Euclidean scalar product. In order to
see a few interesting duality relations we define the polynomials
$$P_j(x)=\frac{P(x)}{x-a_j}=\sum_{k=0}^{n-1}S_k^{(j)}x^{n-k-1},\,\,
j=1,\dots,n$$ From the property $P_j(a_i)=P'(a_i)\delta_{ij}$ we deduce that
the vectors $V_j=(S_0^{(j)},\dots,S_{n-1}^{(j)})$ and
$U_i=(a_i^{n-1},\dots,1)$ are bi-orthogonal, i.e.
$$(V_j,U_i)=P'(a_i)\delta_{ij},\quad i,j=1,\dots,n \eqno{\rm A.2}$$ showing
that $V_j$ and $U_i$ are dual each other. By construction $V_j$ does not
depend on $a_j$.

 It is easily seen that the Vandermonde determinant has two
dual   equivalent definitions $$V_n(a_1,\dots,a_n)=\left|\begin{array}{cccc}
1&1&\dots&1\\ a_1&a_2&\dots&a_n\\ \dots&\dots&\dots&\dots\\
a_1^{n-1}&a_2^{n-1}&\dots&a_n^{n-1}
\end{array}\right|=\left|\begin{array}{cccc} 1&1&\dots&1\\
S_1^{(1)}&S_1^{(2)}&\dots&S_1^{(n)}\\ \dots&\dots&\dots&\dots\\
S_{n-1}^{(1)}&S_{n-1}^{(2)}&\dots&S_{n-1}^{(n)}

\end{array}\right|\eqno {\rm A.3}$$

Let $V_{n-1}^{(j)}$ be the determinant obtained by removing the $j$th column and the last row in $V_n$, then the following relations hold
$$V_{n-1}^{(j)}=\prod_{\stackrel{1\le k<l\le n}{k\ne j\ne l}}(a_l-a_k)$$
$$\prod_{j=1}^n\, V_{n-1}^{(j)}=(V_n)^{n-2}\eqno {\rm A.3}$$
$$\frac{V_n}{V_{n-1}^{(j)}}=(-1)^{n-j}P'(a_j),\quad j=1,\dots, n \eqno{\rm A.4}$$
i.e.  $V_{n-1}^{(j)}$ is the Vandermonde determinant of the indeterminates $a_1,\dots,a_n$, but $a_j$.

We give now the most general form of direct and dual Jacobi identities. 
By replacing the last row of the first form of $V_n$ by the row
$(a_1^k,\dots,a_n^k)$ and expanding over this row we find the identities, see
e.g. \cite{BT} $$\sum_{i=1}^n
(-1)^{n-1}a_i^kV_{n-1}^{(i)}=\left\{\begin{array}{ll} 0& k=0,1,\dots,n-2\\
V_n& k=n-1\\ V_n\sum_{i=1}^n a_i&k=n\\
\end{array}\right.$$
By replacing the last row of the dual form of $V_n$ by $S_k^{(1)},\dots,S_k^{(n)}$ we find the dual Jacobi identity
$$\sum_{i=1}^n (-1)^{n-1} S_k^{(i)}V_{n-1}^{(i)}=\left\{\begin{array}{ll}
0&k=0,\dots, n-2\\
V_n&k=n-1\end{array}
\right. \eqno {\rm A.5}$$
The above formula is a consequence of a more general result. Let $A_{ij}$ be the minor of the $(i,j)$ element of the determinant $V_n$, i.e.
$$A_{ij}=\left|\begin{array}{cccccc}
1&\dots&1&1&\dots&1\\
a_1&\dots&a_{j-1}&a_{j+1}&\dots&a_n\\
\cdot&\cdots&\cdot&\cdot&\cdots&\cdot\\
a_1^{i-1}&\dots&a_{j-1}^{i-1}&a_{j+1}^{i-1}&\dots&a_n^{i-1}\\
\\
a_1^{i+1}&\dots&a_{j-1}^{i+1}&a_{j+1}^{i+1}&\dots&a_n^{i+1}\\
\cdot&\cdots&\cdot&\cdot&\cdots&\cdot\\
a_1^{n-1}&\dots&a_{j-1}^{n-1}&a_{j+1}^{n-1}&\dots&a_n^{n-1}\\
\end{array}\right|$$
From A.2 we get 
$$a_i^{n-1}=-\sum_{k=1}^{n-1}S_k^{(j)}a_i^{n-k-1}\,,\quad i=1,\dots,n$$
and substitute it in the last row of $A_{ij}$. Afterwards we multiply the first row by $S_{n-1}^{(j)}$, the second by   $S_{n-2}^{(j)}$, the $i$th one 
 $S_{n-i}^{(j)}$, etc., and add them to the last row and we get

$$A_{ij}=\left|\begin{array}{cccccc}
1&\dots&1&1&\dots&1\\
a_1&\dots&a_{j-1}&a_{j+1}&\dots&a_n\\
\cdot&\cdots&\cdot&\cdot&\cdots&\cdot\\
a_1^{i-1}&\dots&a_{j-1}^{i-1}&a_{j+1}^{i-1}&\dots&a_n^{i-1}\\
a_1^{i+1}&\dots&a_{j-1}^{i+1}&a_{j+1}^{i+1}&\dots&a_n^{i+1}\\
\cdot&\cdots&\cdot&\cdot&\cdots&\cdot\\
-S_{n-i-1}^{(j)}a_1^i&\dots&-S_{n-i-1}^{(j)}a_{j-1}^i&
-S_{n-i-1}^{(j)}a_{j+1}^i  &\dots&-S_{n-i-1}^{(j)}a_{n}^i\\
\end{array}\right|$$
$$=(-1)^{n-i-1}S_{n-i-1}^{(j)}V_{n-1}^{(j)}(a_1,\dots,a_{j-1},a_{j+1},\dots,a_n)$$
where $V_{n-1}^{(j)}$ is the determinant obtained from $V_n$ by deleting the last row and the $j$-column. 

For obtaining a dual result we denote by $B_{ij}$ the corresponding minor obtained from the dual form of $V_n$. Like the preceding case we use the relation
$$ S_{n-1}^{(l)}=-\sum_{k=0}^{n-2}S_k^{(l)}a_j^{n-k-1},\quad j=1,\dots,n$$
and substitute it in the last row of $B_{ij}$. Multiplying the first row by $a_j^{n-1}$, the second by  $a_j^{n-2}$, etc., and adding to the last row we find

$$B_{ij}=\left|\begin{array}{cccccc}
1&\cdots&1&1&\cdots&1\\
S_1^{(1)}&\cdots&S_1^{(j-1)}&S_1^{(j+1)}&\cdots&S_1^{(n)}\\
\cdot&\cdots&\cdot&\cdot&\cdots&\cdot\\
S_{i-1}^{(1)}&\cdots&S_{i-1}^{(j-1)}&S_{i-1}^{(j+1)}&\cdots&S_{i-1}^{(n)}\\
S_{i+1}^{(1)}&\cdots&S_{i+1}^{(j-1)}&S_{i+1}^{(j+1)}&\cdots&S_{i+1}^{(n)}\\
\cdot&\cdots&\cdot&\cdot&\cdots&\cdot\\
-a_j^{n-i-1}S_i^{(1)}&\cdots&-a_j^{n-i-1}S_i^{(j-1)}&-a_j^{n-i-1}S_i^{(j+1)}&\cdots&-a_j^{n-i-1}S_i^{(n)}\\
\end{array}\right|$$

$$=(-a_j)^{n-i-1}V_{n-1}^{(j)}(a_1,\dots,a_{j-1},a_{j+1},\dots,a_n)$$

\end{document}